\documentclass[aps,fleqn,twocolumn,showpacs]{revtex4}

\usepackage{graphicx}
\usepackage{amsmath} 

\def\At{{A_t}}
\def\phit{{\phi_t}}
\def\Omegat{{\Omega_t}}
\def\Gammat{{\Gamma_t}}
\def\vcom{v_\mathrm{com}}
\def\vcom{v_{\scriptscriptstyle\mathrm{CM}}}
\def\com{\textsc{cm}}

\begin{document}

\title{ Ionization of clusters in intense laser pulses 
  through collective electron dynamics}
\author{Ulf Saalmann}
\author{Jan-Michael Rost}
\affiliation{Max-Planck-Institut f{\"u}r Physik komplexer Systeme,
  N{\"o}thnitzer Str.\ 38, 01187 Dresden, Germany}
\date{\today}

\begin{abstract}\noindent
  The motion of electrons and ions in medium-sized rare
  gas clusters ($\sim$\,1000\,atoms) exposed to intense laser
  pulses is studied microscopically by means of classical
  molecular dynamics using a hierarchical tree code. 
  Pulse parameters for optimum ionization are found to be
  wavelength dependent.
  This resonant behavior is traced back to a collective electron
  oscillation inside the charged cluster. 
  It is shown that this dynamics can be well described by a driven
  and damped harmonic oscillator allowing for a clear
  discrimination against other energy absorption mechanisms. 
\end{abstract}


\maketitle

The interaction of of intense laser radiation with clusters has
been of continuing interest
\cite{mcth+94,diti+97,ledo+98,dizw+99,zwdi+99,kosc+99,po01},
pushed by prominent findings as
the emission of keV-photons \cite{mcth+94},
highly charged ions \cite{diti+97,ledo+98},
or fast fragments \cite{dizw+99}.
Basically, all these phenomena are caused by the exceedingly
effective absorption of energy from the laser field into the cluster.
This enhanced absorption (when compared to atoms or bulk matter)
is possible due the 
initially \emph{solid-like atomic density in the cluster} 
in combination with the \emph{rapid expansion of the cluster}
on a femtosecond time scale, i.\,e.\ typically during the laser pulse.
More detailed insight into the mechanism of energy absorption
can be gained by pulse length variation.
Using pulses of equal energy, i.\,e.\ longer pulses have lower
intensities, it has been found that there is an optimum pulse
length with maximum absorption \cite{zwdi+99,kosc+99}. This holds true
for small rare gas and metal clusters (of the order of some 10
atoms) as well as for large clusters, although the underlying
reason for maximum absorption can be very different and is in
fact an issue of current debate. 
For small rare gas clusters under laser pulses of peak
intensities in the tunneling regime ($I\agt10^{15}$\,W/cm$^2$)
such an optimum could be attributed to the mechanism of enhanced
ionization \cite{scro0203} known from molecules \cite{zuba95seiv+95}. 
For small metal
clusters exposed to similar laser pulses the existence of
optimal absorption was interpreted as a plasmon resonance
phenomenon \cite{kosc+99} in analogy to the well known dipole
resonance of the valence electrons in perturbative
photo-absorption \cite{br93}
or low-intensity laser pulses \cite{sure00}. 
For large clusters (of more than $10^{5}$ atoms) it has been
proposed that strong laser pulses create a nanoplasma inside the
cluster \cite{dido+96,mimc+01}. 
In the course of the expansion of the cluster the
electron density and consequently the plasma frequency decreases
resulting in strong energy absorption at resonance  with the
laser.

Separating the different mechanisms from each other requires
specific and clear signatures for each process which are
difficult to identify in a multi particle system such as a
cluster. Surprisingly, dipole resonant absorption dynamics in a
cluster can be very well characterized by a simple driven damped
harmonic oscillator, which describes the dipole response of the
electrons inside the cluster.  

In the following we will demonstrate the validity of this simple
description with  full dynamical microscopic calculations for
Xenon clusters ($\sim10^2$\ldots10$^3$ atoms) in strong optical
laser pulses 
(wavelengths $\lambda=520\ldots1170$\,nm,
intensities $I\sim10^{14}$\ldots10$^{16}$\,W/cm$^2$,
pulse lengths $T\sim10\ldots1000$\,fs). 
Our approach is similar to those used before for intense
laser-cluster interaction \cite{rosc+97}. However, we have been
forced to use a completely new propagation scheme, namely a
hierarchical tree code \cite{pfgi96}, to handle of the order of
$10^4$ charged particles 
($\sim$\,1000 ions and $\sim$\,8000 electrons) 
with their mutual interactions.  
Originally developed for gravitational $N$-body problems in
cosmology \cite{bahu86}, the hierarchical tree code allows us to
follow the dynamics of all charged particles over a few hundred
femtoseconds with typical time steps of attoseconds. 

Atoms are initially arranged in so-called Mackay icosahedra
\cite{ho79}  
known to be optimal structures of Lennard-Jones clusters.
We distinguish between free and bound electrons, whereby the latter
ones are not treated explicitly.
The condition for creation of an electron is that none of the other electrons
has a negative binding energy 
to the ion under consideration.
In this case a new electron is ``born'' at the position of that ion
with a kinetic energy to satisfy the ionization potential.
The charge of the ion is increased by one.   
Free electrons as well as ions are described classically
as charged particles in an oscillating field
interacting via a softened Coulomb interaction $W$.
This is defined for a pair of particles with charges $q_1$ and
$q_2$ and a distance of $r_{12}$ as
$W=q_1\,q_2/\sqrt{{r_{12}}^2+1}$.

\begin{figure}[t]
  \centering
  \includegraphics{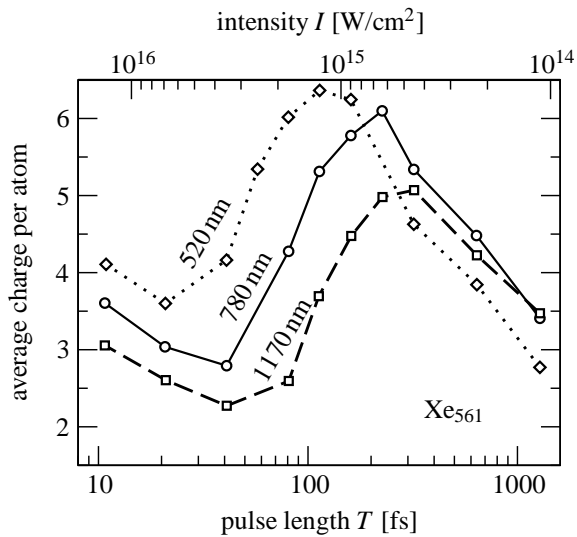}
  \caption{Average charge per atom from Xe$_{561}$ 
    clusters after laser impact 
    according to Eq.~(\ref{eq:lapu})
    as a function of the pulse length $T$ or 
    the peak intensity $I$, respectively, 
    for three different laser wavelengths $\lambda$.
    The energy of the pulse, 
    i.\,e.\  the product $I\times T$, 
    is kept constant.}
  \label{fig:pule}
\end{figure}

First of all, we present the calculated pulse length dependence
for three different laser wavelengths $\lambda$.
Figure~\ref{fig:pule} shows the final averaged charge state per
atom from Xe$_{561}$ clusters after laser impact
as a function of the pulse length $T$ or 
the peak intensity $I$, respectively.
The applied laser pulse was linearly polarized with a field

\begin{equation}
  \label{eq:lapu}
  F(t) =
  \left\{\begin{array}{ll}
      \sqrt{I/I_0}\cos^2
      \left(\frac{\pi}{2}\frac{t}{T}\right)
      \cos(\omega t)
      & \mbox{for }|t|<T
      \\ [1ex] 0 & \mbox{elsewhere}
    \end{array}\right.
\end{equation}
where $I_0=3.51{\cdot}10^{16}$\,W/cm$^2$.
In order to keep the energy of the different pulses constant
we fixed the product of intensity and pulse length
$I\times T=I_0\times4$\,fs.
For short pulses ($T\,{\alt}\,30$\,fs), where the cluster atoms have
not enough time to react on the charging, the final charge state
decreases with an increasing pulse length due to the lower intensity.
For longer pulses ($T\,{\agt}\,50$\,fs), however, the final charge
state increases despite the intensities become smaller. 
This can only be understood if one considers the expansion of
the cluster, for a detailed explanation  see below.
Finally, for very long pulses ($T\,{\agt}\,400$\,fs) the cluster is
already completely fragmented before the laser pulse reaches its peak
intensity rendering the ionization similar compared to the case of single
atoms.  
Qualitatively, this behavior is the same for all three frequencies (Fig.~\ref{fig:pule}).
However, the shift of the optimal pulse length  towards longer pulses for longer wavelengths
is characteristic for a resonant ionization mechanism.

\begin{figure}[b]
  \centering
  \includegraphics{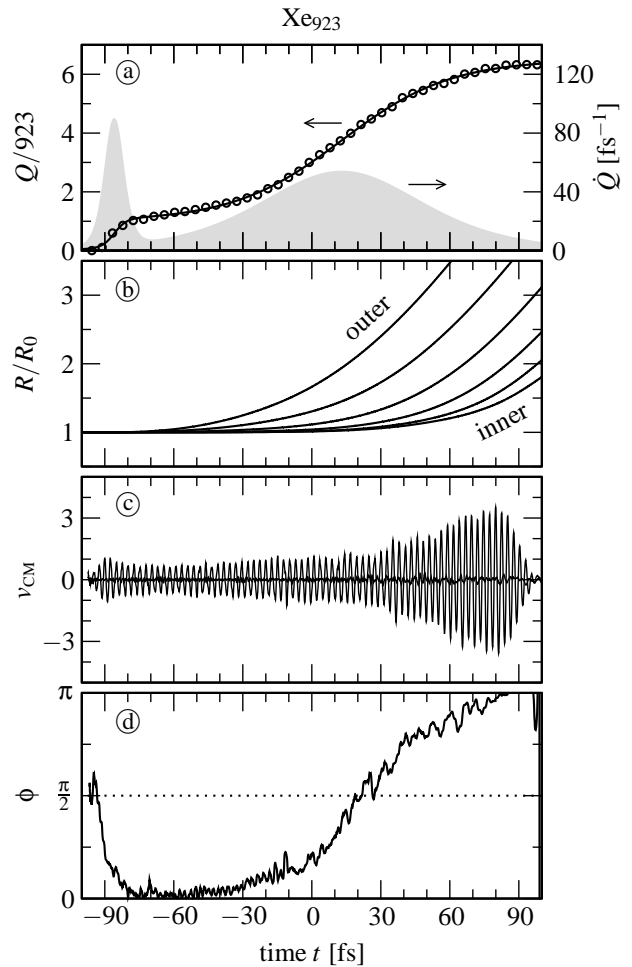}
  \caption{Dynamics of Xe$_{923}$ in a strong laser pulse 
    ($\lambda=780$\,nm, $I=9\cdot10^{14}$\,W/cm$^2$,
    rise and fall time 20\,fs, plateau for $t=-80\ldots+80$\,fs).
    All quantities are shown as a function of time $t$.
    a: Average charge per atom (\textsl{circles and fit from
    Eq.~(\ref{eq:tsff}), left axis})
    and corresponding rate (\textsl{gray filled line, right axis}).
    b: Radii $R$ of all cluster shells in units of their
    initial radii $R_0$.
    c: Centre-of-mass velocity $\vcom$ 
    of the electronic cloud inside the
    cluster volume. Note, that the oscillations are spatially along  
    the linear polarization of the laser, whereas the electron
    velocity perpendicular to the laser polarization  is very
    small and  hardly to see in the figure. 
    d: Phase shift $\phit$ of the collective oscillation in laser
    direction with respect to the driving laser, see text.}
  \label{fig:time}
\end{figure}%
To gain insight into the mechanism of ionization we discuss the
dynamics of a Xe$_{923}$ cluster in a laser pulse with 
$I=9\cdot10^{14}$\,W/cm$^2$, cf.\ Fig.~\ref{fig:time}. 
The pulse has short rise and fall times of 20\,fs and a long
plateau of 160\,fs in order to eliminate effects from the time
dependence of the laser pulse itself. 
As can be clearly seen in Fig.~\ref{fig:time}a, 
the cluster ionization occurs in two steps 
and the total cluster charge $Q$ 
(sum of total ionic charge $Q_\mathrm{ion}$ and 
charge of those electrons which are inside the cluster volume)
as a function of time $t$ is well represented by
\begin{equation}
  \label{eq:tsff}  
  Q(t)=\sum_{i=1,2}\frac{Q_i}{1+\exp(-(t-t_i)/\delta t_i)}.
\end{equation}
In the 1st step during the rising of the pulse 
($t_1\pm\delta t_1=-86$\,fs\,$\pm$\,3\,fs)
electrons are emitted mainly due to field ionization.
This process slows down, however, already before the plateau
intensity is reached due to the increasing
space charge (cluster charge $Q\approx1000$ at $t=-80$\,fs).
Because of this attractive space charge, 
one may distinguish between inner ionization,
which accounts for excitation from localized electrons to
quasi-free electrons moving inside the cluster volume,
and outer ionization, which corresponds to the final escape of
the quasi-free electrons from the cluster into the continuum.
Note, that the space charge can hold an appreciable number of
quasi-free electrons which engage into collective motion
discussed below, 
in contrast to  the ionization dynamics of molecules 
or small clusters  with almost no quasi-free electrons
\cite{scro0203}. 
The charging up of the cluster leads to an expansion
as can be seen for $t\agt-60$\,fs in Fig.~\ref{fig:time}b.
During the expansion a 2nd ionization step occurs
which lasts for a much longer time and leads to an increase of the
average ionic charge from about 1 to more than 6. 
During this time, the quasi-free electrons in the cluster
are driven collectively back and forth
along the polarization direction of the laser which is
evident from their centre-of-mass (\com) velocity
$\vcom$ shown in Fig.~\ref{fig:time}c.

This oscillation can be modelled by a driven
and damped classical harmonic oscillator
\begin{equation}
  \label{eq:ddho}
  \ddot{X}(t)+2\Gammat \dot{X}(t)+\Omegat^2 X(t)=F(t)
\end{equation}
with $X(t)$ the \com\ position of the electron cloud,
$F(t)$ the driving laser amplitude, 
and $\Omegat$ and $\Gammat$ the eigenfrequency and damping
rate, respectively, which are determined by the cluster. 
The index $t$ indicates 
that due to ionization and expansion of the cluster, both,
$\Omegat$ and $\Gammat$, may depend parametrically on time. 
Under the assumption of a spherical, uniformly charged
cluster with total ionic charge $Q_\mathrm{ion}$
and radius $R$ the potential inside the cluster is
harmonic with an eigenfrequency 
$\Omegat=\sqrt{Q_\mathrm{ion}(t)/R(t)^3}$.
The damping is caused by both, internal heating of the
quasi-free electrons in the cloud and energy transfer to bound
electrons. 
These two effects are responsible for outer and inner
ionization, respectively. 

For  periodic driving $F(t)=F_{0}\cos(\omega t)$ 
the dynamics is given by
$X(t)=\At\cos(\omega t-\phit)$
with \cite{lali94}
\begin{subequations}
  \label{eq:amph}
  \begin{eqnarray}
    \At &=&
    F_{0}\big/\sqrt{\big(\Omegat^2-\omega^2\big)^2+(2\Gammat\omega)^2},
    \label{eq:ampl} 
   \\
    \phit &=& 
    \arctan\left(2\Gammat\omega\big/(\Omegat^2-\omega^2)\right).
    \label{eq:phas}
  \end{eqnarray}
\end{subequations}
The energy balance of the dynamics (\ref{eq:amph}) 
is characterized, on one hand, by energy loss
$E_\mathrm{loss}$ due to the damping
and, on the other hand, by energy gain $E_\mathrm{gain}$ 
from the external laser field. 
The cycle-averaged energy transfer rates read
\begin{eqnarray}
  \langle\dot{E}\rangle &=&  
  \langle\dot{E}_\mathrm{loss}\rangle+\langle\dot{E}_\mathrm{gain}\rangle
  \nonumber\\
  &=& -\Gammat\,\At^2\,\omega^2+\frac{1}{2}F_{0}\,\At\,\omega\,\sin\phit.
  \label{eq:enba}  
\end{eqnarray}
Obviously and well known \cite{lali94}, 
maximum $\langle\dot{E}_\mathrm{gain}\rangle$ or
optimal heating requires $\phit=\pi/2$,
i.\,e.\ resonant behaviour $\Omegat=\omega$.
As shown in Fig.~\ref{fig:time}d, the phase shift 
$\phit$ changes in time from 0 to $\pi$,
thereby passing the resonance $\phit=\pi/2$.
This change is directly connected with the increased
ionization of the cluster, cf.\  Fig.~\ref{fig:time}a.
In particular, the resonance time $t_\mathrm{res}$
coincides with the time $t_2$ of maximal ionization rate, 
cf.\ Eq.~(\ref{eq:tsff}).
This applies to other laser wavelengths as well
(For $\lambda=1170$\,nm $t_2$ is somewhat smaller
due to the early laser switch-off at $t=80$\,fs.):
\smallskip\begin{ruledtabular}
  \begin{tabular}{clcccc}
    & $\lambda_\mathrm{laser}$
    & 520\,nm & 780\,nm & 1170\,nm \\ \hline
    & $t_2\pm\delta t_2$
    & ($-$24$\pm$13)\,fs & (13$\pm$24)\,fs & (40$\pm$26)\,fs \\
    & $t_\mathrm{res}$
    & $-$22\,fs & 19\,fs & 72\,fs \\
  \end{tabular}
\end{ruledtabular}\bigskip\noindent
\begin{figure}[t]
  \centering
  \includegraphics{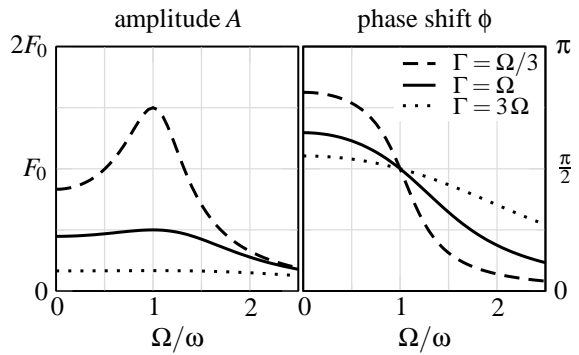}
  \caption{Driven and damped harmonic oscillator:
    amplitude $A$ and phase shift $\phi$ as a function of
  the ratio of eigenfrequency $\Omega$ and driving frequency 
  $\omega$ according to Eqs.~(\ref{eq:amph})
  for different damping strengths $\Gamma$.}
 \label{fig:ampa}
\end{figure}%
Passing through the resonance is not necessarily connected with
large amplitude oscillations, if the damping strength $\Gammat$ is
comparable to or larger than the eigenfrequency $\Omegat$, 
see Fig.~\ref{fig:ampa}.
Otherwise, the phase shift $\phit$ at resonance is independent
of the damping (Fig.~\ref{fig:ampa}).

\begin{figure}[t]
  \centering
  \includegraphics{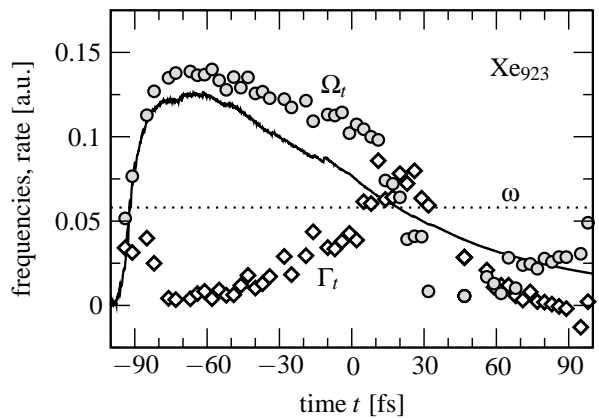}
  \caption{Parameters of the harmonic oscillator
    model~(\ref{eq:ddho}) as calculated
    from the Xe$_{923}$ dynamics in Fig.~\ref{fig:time}.
    \textsl{Solid line:} eigenfrequency for a spherical,
    uniformly charged cluster $\Omegat=\sqrt{Q_\mathrm{ion}(t)/R(t)^3}$.
    \textsl{Circles:}  eigenfrequency $\Omegat$~(\ref{eq:omeg}).
    \textsl{Diamonds:} damping rate $\Gammat$~(\ref{eq:gamm}).
    \textsl{Dotted line:} laser frequency $\omega$.
}
 \label{fig:omga}
\end{figure}

In order to validate the applicability of the driven and damped
harmonic oscillator model of Eq.~(\ref{eq:ddho}) 
we use amplitude $\At$ and phase shift $\phit$
of the collective oscillation from our results 
of Fig.~\ref{fig:time} to determine frequency and damping rate
according to
\begin{subequations}
  \label{eq:omga}
  \begin{eqnarray}
    \Omegat^2 &=&
    \omega^2 + (F_{0}/\At)\cos\phit,
    \label{eq:omeg} 
   \\
    \Gammat &=& 
    \big(F_{0}/(2\At\omega)\big)\sin\phit.
    \label{eq:gamm}
  \end{eqnarray}
\end{subequations}
Figure~\ref{fig:omga} shows these parameters as a function of
time $t$ for the same system as in Fig.~\ref{fig:time}.
The calculated eigenfrequency $\Omegat$ 
(gray circles in Fig.~\ref{fig:omga})  
closely resembles the frequency of a uniformly charged sphere  
(solid line).
Moreover, both frequencies match the laser frequency $\omega$ at
the same time $t\approx20$\,fs providing additional support for
the collective oscillator model. 
Along with the decrease of the eigenfrequency $\Omegat$ 
the damping term $\Gammat$ (white diamonds in Fig.~\ref{fig:omga})
rises for times $t\approx-60\ldots30$\,fs. 
This accounts for energy transfer to deeper and deeper bound
electrons,  
which does not occur in other theoretical studies where also
resonant behaviour was discussed \cite{lajo99,paco+01}. 
Either inner ionization was not considered \cite{lajo99}
or deuterium clusters composed of single electron atoms
were discussed \cite{paco+01}.
However, in order to understand the experimentally observed high
charge states \cite{diti+97,ledo+98,zwdi+99,kosc+99}
it is of utmost importance to take this into consideration.
It is just this continuous cycle of effective heating and
induced inner ionization at the resonance which causes the
increased ionization rate and the high charge states of the
fragments.
At this point it is interesting to note, that the damping almost
completely compensates the heating as can be deduced from the
almost negligible increase of the amplitude of $\vcom$
before reaching the resonance (Fig.~\ref{fig:time}c).
Once the cluster has passed the resonance, however, 
the damping falls off rapidly 
($t\agt30$\,fs in Fig.~\ref{fig:omga}).
Obviously, the oscillating electron cloud becomes unable to drive
further inner ionization. 
This implies a weaker damping and thus an increase 
of the oscillation amplitude, cf.\ Fig.~\ref{fig:time}c.

In summary, we have shown that the electron emission in
medium-sized rare gas clusters ($\sim$\,10$^3$\,atoms) 
is enhanced by resonant energy absorption  in agreement with
experimental data \cite{zwdi+99,kosc+99}.
(The metallic nature of the clusters used in
\cite{kosc+99}
should be of minor importance for the creation of the high
charge states $\agt8$
since the delocalized valence electrons are emitted early in the
pulse.) 
Our microscopic calculations of the motion of ions and electrons
using a hierarchical tree code reveal a laser-driven collective 
oscillation of the cloud of quasi-free electrons which are held
back inside the cluster volume by the space charge of the cluster.
The eigenfrequency of this oscillation is determined by
charge and size of the cluster.
Electron emission and cluster expansion change these quantities in time
and eventually enable the matching of eigenfrequency $\Omegat$
and laser frequency $\omega$ during the pulse. 
This resonance allows for an effective energy transfer to the
collective motion.  
The changing phase shift between driving field and driven
electron cloud clearly indicates the different stages of
energy absorption of the cluster electrons from the laser field.
The fact that the collective electron dynamics can be well described by 
a simple damped harmonic oscillator helps to clarify the nature of this dynamics and
provides a clear signature of this type of collective dynamics. 
We expect that the other two mechanisms for effective energy
absorption, namely enhanced ionization and nanoplasma excitation
will have an almost vanishing amplitude for the
electronic \com\ velocity.
Small clusters, which exhibit enhanced ionization, produce only
a few quasi-free electrons which cannot create a sizable
\com-velocity amplitude.
In large clusters, however, where a nanoplasma is formed,
electrons are heated resonantly at their critical density
leaving their \com\ position at rest.

\end{document}